\documentclass[10.5pt]{article} 

\usepackage[margin=0.7in]{geometry} 
\usepackage{amsmath,amssymb}
\usepackage{bm}
\usepackage[dvipdfmx]{graphicx}
\usepackage{verbatim}
\usepackage{wrapfig}
\usepackage{ascmac}
\usepackage{makeidx}
\usepackage{fancybox} 
\usepackage{txfonts} 
\usepackage{multicol} 
\usepackage{float} 
\usepackage{booktabs,caption}
\usepackage{cite}
\usepackage{url}
\usepackage[flushleft]{threeparttable}
\usepackage{titlesec}
\usepackage[table]{xcolor}
\usepackage{booktabs}
\usepackage{colortbl}
\definecolor{mygray}{gray}{0.9}

\usepackage{lineno}

\newcommand*\patchAmsMathEnvironmentForLineno[1]{
	\expandafter\let\csname old#1\expandafter\endcsname\csname #1\endcsname
	\expandafter\let\csname oldend#1\expandafter\endcsname\csname end#1\endcsname
	\renewenvironment{#1}
	{\linenomath\csname old#1\endcsname}
	{\csname oldend#1\endcsname\endlinenomath}}
\newcommand*\patchBothAmsMathEnvironmentsForLineno[1]{
	\patchAmsMathEnvironmentForLineno{#1}
	\patchAmsMathEnvironmentForLineno{#1*}}
\AtBeginDocument{
	\patchBothAmsMathEnvironmentsForLineno{equation}
	\patchBothAmsMathEnvironmentsForLineno{align}
	\patchBothAmsMathEnvironmentsForLineno{flalign}
	\patchBothAmsMathEnvironmentsForLineno{alignat}
	\patchBothAmsMathEnvironmentsForLineno{gather}
	\patchBothAmsMathEnvironmentsForLineno{multline}
}

\title{Statistical Inference of the Matthews Correlation Coefficient for Multiclass Classification}

\author{Jun Tamura\thanks{Graduate School of Medicine, Yokohama City University, Japan}, \and
	Yuki Itaya\thanks{Graduate School of Science and Technology, Keio University, Japan},\and
	Kenichi Hayashi\thanks{Department of Mathematics, Keio University, Japan}, \and
	Kouji Yamamoto\thanks{Department of Biostatistics, School of Medicine, Yokohama City University, Japan}}

\begin{document}
	\maketitle
	\begin{abstract}
		Classification problems are essential statistical tasks that form the foundation of decision-making across various fields, including patient prognosis and treatment strategies for critical conditions. Consequently, evaluating the performance of classification models is of significant importance, and numerous evaluation metrics have been proposed. Among these, the Matthews correlation coefficient (MCC), also known as the phi coefficient, is widely recognized as a reliable metric that provides balanced measurements even in the presence of class imbalance. However, with the increasing prevalence of multiclass classification problems involving three or more classes, macro-averaged and micro-averaged extensions of MCC have been employed, despite a lack of clear definitions or established references for these extensions. In the present study, we propose a formal framework for MCC tailored to multiclass classification problems using macro-averaged and micro-averaged approaches. Moreover, discussions on the use of these extended MCCs for multiclass problems often rely solely on point estimates, potentially overlooking the statistical significance and reliability of the results. To address this gap, we introduce several methods for constructing asymptotic confidence intervals for the proposed metrics. Furthermore, we extend these methods to include the construction of asymptotic confidence intervals for differences in the proposed metrics, specifically for paired study designs. The utility of our methods is evaluated through comprehensive simulations and real-world data analyses.

	\end{abstract}
	
	
	\vspace{0.5cm}
	
	\noindent\textbf{Correspondence:} \\
	Author Name: Jun Tamura \\
	Institution: Graduate School of Medicine, Yokohama City University, Japan \\
	Address: 3-9 Fukuura, Kanazawa-ku, Yokohama 236-0004, Japan\\
	Email: \texttt{light.130728ff@outlook.jp} \\
	
\noindent\textbf{Funding information}\\
	Funder: 
	\begin{itemize}
		\item Japan Society for the Promotion of Science
		\begin{itemize}
			\item Grant Number 1: 21K117907
			\item 	Grant Number 2: 23K11013
		\end{itemize}
		\item Japan Science and Technology Agency (JST) Support for Pioneering Research Initiated by the Next Generation
		\begin{itemize}
			\item Grant Number: JPMJSP2179
		\end{itemize}
	\end{itemize}

	\section{Introduction}
	In the field of medicine, evaluating the predictive capabilities of models is of paramount importance. Predictive models serve various purposes, including definitive diagnosis, disease screening, and prognosis prediction, among others. Metrics such as sensitivity, specificity, positive predictive value, and negative predictive value are widely employed to assess the performance of binary predictive models. Furthermore, the advent of big data, which facilitates  easy data collection, has accelerated the development of classifiers across diverse fields. In addition to commonly used metrics for model performance evaluation, measures such as the $\text{F}_1$ score and the Matthews correlation coefficient (MCC) have been proposed. Notably, MCC is recognized for providing a balanced assessment even in the presence of class imbalances, making it a robust and reliable metric.
	
	MCC is derived by adapting Pearson's correlation coefficient to binary data, yielding values ranging from $-1$ to $1$. An MCC of $1$ indicates perfect agreement between predicted and true classifications, $0$ signifies random predictions, and $-1$ reflects complete disagreement between predicted and true classifications. MCC is particularly reliable for evaluating classifier performance because it provides highly informative and robust insights without being overly influenced by individual elements. While there is no consensus on the most suitable metric for specific scenarios, recent studies have emphasized MCC's reliability and its utility across various fields.
	
	Studies advocating for the use of MCC, such as those by Chicco et al.\cite{ch20,ch21}, have highlighted several advantages of MCC through comparative analyses with metrics like accuracy, balanced accuracy, and the $\text{F}_1$ score. Their findings suggest prioritizing MCC in most studies, particularly for achieving values close to $1$, as it comprehensively accounts for all relevant factors and provides high reliability. Similarly, Yao et al.\cite{yao} recommend replacing the $\text{F}_1$ score with the unbiased MCC, especially for imbalanced datasets, as it overcomes the inherent bias of the $\text{F}_1$ score and enables more accurate performance evaluations.
	
	Moreover, Itaya et al.\cite{ita} derived asymptotic confidence intervals for MCC in binary classifications using the delta method, addressing the uncertainty of point estimates. With the increasing prevalence of multi-class classification problems, the importance of MCC in such contexts has grown. However, the application of MCC in multi-class classification remains underdeveloped, with clear definitions and methodologies still lacking. Many existing metrics focus primarily on binary classification, often leading to suboptimal assessments when applied to multi-class scenarios by reducing the problem to binary cases\cite{al,fb}. To address this issue, various measures, such as macro-average and micro-average $\text{F}_1$ scores, have been extended for multi-class classification. Takahashi et al.\cite{taka21,taka23} proposed methods for conducting statistical inference using micro-average and macro-average $\text{F}_1$ scores. Similarly, extensions of MCC, such as macro-average MCC and micro-average MCC, have been widely adopted by practitioners for multi-class classification\cite{tal,li,sam,ahu,tra,alle}.
	
	For example, Allenbrand et al.\cite{alle} proposed an image classification approach to assess the severity of COVID-19 infection from lung CT scans, evaluating its performance using macro-average MCC point estimates alongside other metrics. However, these extensions often lack clear definitions, and statistical inference methods for them remain underexplored.
	
	Additionally, Gorodkin\cite{goro} proposed the $R_k$ statistic as an extension of MCC for binary classification, yet statistical inference for this metric is also insufficiently developed.
	
	The present study aims to formalize the MCC used by practitioners for multi-class classification and derive its asymptotic distribution using the central limit theorem and the delta method. Furthermore, it explores methods for constructing confidence intervals for the difference between formulated measures in paired designs.
	
	The remainder of this paper is organized as follows: Section 2 defines MCC for binary classification and extends it to multi-class classification using macro-average and micro-average approaches. Section 3 describes two methods for deriving asymptotic confidence intervals for individual MCC values . Section 4 elaborates on methods for obtaining confidence intervals for the difference between formulated measures in paired designs. Sections 5 and 6 validate the proposed methods through simulation studies and real-data analysis, respectively. Finally, the paper concludes with a summary of the findings and a discussion of future research directions.

	\section{Definition of Matthews Correlation Coefficient}
	In this section, we introduce the notation and definition of a binary MCC (biM), followed by the notation and definition of multi-class MCCs. Consider an $r \times r$ table, as depicted in Table 1, representing the outcomes of a prediction and the true class. In this context, each subject is assumed to be a random sample from the population distribution of interest, involving classification into $r$ distinct classes: the predicted class $X$ and the true class $Y$. Define $\pi_{ij} = P(X=i, Y=j)$ for $i,j= {1,\ldots,r}$ as the probability that $X$ and $Y$ belong to the cell in row $i$ and column $j$:
	\begin{eqnarray*}
		\text{True Positive (TP)}: \mathrm{TP}_i&=&P(X=i, Y=i)=\pi_{ii},\\
		\text{False Positive (FP)}: \mathrm{FP}_i&=&P(X=i, Y\neq i)=\sum_{\substack{j=1 \\ j\neq i}}^r \pi_{ij},\\
		\text{False Negative (FN)}: \mathrm{FN}_i&=&P(X\neq i, Y=i)=\sum_{\substack{j=1 \\ j\neq i}}^r \pi_{ji},\\
		\text{True Negative (TN)}: \mathrm{TN}_i&=&P(X\neq i, Y\neq i)=1-\mathrm{TP}_i-\mathrm{FP}_i-\mathrm{FN}_i.
	\end{eqnarray*}
	\begin{table}[hbtp]
		\centering
		\caption{Prediction $(X)$ and true classification $(Y)$}
		\centering
		\begin{tabular}{lccccc}
			\hline
			&& \multicolumn{4}{c}{$\text{True Classification:}Y$} \\
			&& 1  &  2 & $\cdots$ & $r$  \\
			\hline \hline
			& 1 & $\pi_{11}$  &$\pi_{12}$&$\cdots$&$\pi_{1r}$ \\
			& 2 & $\pi_{21}$   &$\pi_{22}$&$\cdots$&$\pi_{2r}$ \\
			$\text{Prediction:}X$& \vdots & \vdots  &\vdots&$\ddots$ &\vdots   \\
			&  $r$ & $\pi_{r1}$   &$\pi_{r2}$&$\cdots$&$\pi_{rr}$\\
			\hline
		\end{tabular}
	\end{table}
	\subsection{Binary MCC}
	When $r=2$, for row variables and column variables having marginal distributions $\{\pi_{+i}\}$ and $\{\pi_{j+}\}$, $\mathrm{biM}$ is expressed as follows:
	
	\begin{gather*}
		\mathrm{biM}=\frac{\mathrm{TP}_1\times \mathrm{TN}_1 - \mathrm{FP}_1 \times \mathrm{FN}_1}{\sqrt{(\mathrm{TP}_1+\mathrm{FP}_1)(\mathrm{TP}_1+\mathrm{FN}_1)(\mathrm{TN}_1+\mathrm{FP}_1)(\mathrm{TN}_1+\mathrm{FN}_1)}}=\frac{\pi_{11}\pi_{22}-\pi_{12}\pi_{21}}{\sqrt{\pi_{+1}\pi_{1+}\pi_{+2}\pi_{2+}}}.
	\end{gather*}
	\subsection{Macro-averaged MCC}
	Macro-averaged MCC (maM) calculates the simple average of $\mathrm{biM}$ for each class, and can be expressed using the following formula\cite{ahu}:
	
	\begin{eqnarray}
		\mathrm{maM}=\frac{1}{r} \sum_{i=1}^r \frac{(\mathrm{TP}_i \times \mathrm{TN}_i - \mathrm{FP}_i \times \mathrm{FN}_i)}{\sqrt{(\mathrm{TP}_i+\mathrm{FP}_i)(\mathrm{TP}_i+\mathrm{FN}_i)(\mathrm{TN}_i+\mathrm{FP}_i)(\mathrm{TN}_i+\mathrm{FN}_i)}}.
	\end{eqnarray}

	\subsection{Micro-averaged MCC}
	There is no universally accepted definition for micro-averaged MCC (miM). In general terms, micro-average pools per-sample classifications across classes and then calculates the overall MCC\cite{alle}. Pooling each cell, we can express it as follows:
	
	\begin{eqnarray*}
		\sum_{i=1}^r \mathrm{TP}_i&=& \sum_{i=1}^r \pi_{ii}, \\
		\sum_{i=1}^r \mathrm{FN}_i &=& 1 - \sum_{i=1}^r \mathrm{TP}_i, \\
		\sum_{i=1}^r \mathrm{FP}_i &=& 1 - \sum_{i=1}^r \mathrm{TP}_i, \\
		\sum_{i=1}^r \mathrm{TN}_i &=& r - 2 + \sum_{i=1}^r \mathrm{TP}_i.
	\end{eqnarray*}
	Then miM can be expressed as follows:
	\begin{eqnarray}
		\mathrm{miM} &=& \frac{(\sum_i \mathrm{TP}_i \times \sum_i \mathrm{TN}_i - \sum_i \mathrm{FP}_i \times \sum_i \mathrm{FN}_i)}{\sqrt{(\sum_i \mathrm{TP}_i + \sum_i \mathrm{FP}_i)(\sum_i \mathrm{TP}_i + \sum_i \mathrm{FN}_i)(\sum_i \mathrm{TN}_i + \sum_i \mathrm{FP}_i)(\sum_i \mathrm{TN}_i + \sum_i \mathrm{FN}_i)}}=\frac{r \sum_i \pi_{ii} - 1}{r - 1}.
	\end{eqnarray}
	However, the essence of \(\mathrm{biM}\) lies in Pearson's correlation coefficient for binary data. Nevertheless, \(\mathrm{miM}\) does not have a lower bound of \(-1\), thus failing to preserve the properties of Pearson's correlation coefficient. Moreover, one of the advantages of MCC is that it provides a reasonable evaluation due to its richness of information, even in situations where other metrics like accuracy and the $\text{F}_1$ score might be overly optimistic. In contrast, \(\mathrm{miM}\) is composed solely of the diagonal elements and has a proportional relationship with accuracy, thereby not preserving the advantages of MCC. Therefore, alternative aggregation methods need to be considered.
	
	Considering a binary classification with \( r = 2 \), since MCC is defined as Pearson's correlation coefficient for binary data, we can express MCC for binary classification as biM:

	\begin{eqnarray*}
		\mathrm{biM} = \frac{\mathrm{Cov}(X,Y)}{\sqrt{\mathrm{Var}(X)\mathrm{Var}(Y)}},
	\end{eqnarray*}
	where \(\mathrm{Cov}(X,Y)\) and \(\mathrm{Var}(X)\), \(\mathrm{Var}(Y)\) can be expressed as
	
	\begin{eqnarray*}
		\mathrm{Cov}(X,Y) &=& \mathrm{{TP}_1} \times \mathrm{{TN}}_1 -  \mathrm{{FP}}_1 \times \mathrm{{FN}}_1, \\
		\mathrm{Var}(X) &=& (\mathrm{{TP}}_1 +  \mathrm{{FP}}_1)(\mathrm{{TN}_1}+ \mathrm{{FN}}_1), \\
		\mathrm{Var}(Y) &=& (\mathrm{{TP}}_1 + \mathrm{{FN}}_1)(\mathrm{{TN}}_1 +  \mathrm{\widehat{FP}}_1).
	\end{eqnarray*}
	Thus, an alternative definition of micro-averaged MCC (\(\mathrm{miM^*}\)) can be expressed in the following form:

	\begin{eqnarray}
		\mathrm{miM^*} &=& \frac{\sum_i (\mathrm{TP}_i \times \mathrm{TN}_i - \mathrm{FP}_i \times \mathrm{FN}_i)}{\sqrt{\sum_i((\mathrm{TP}_i + \mathrm{FP}_i)(\mathrm{TN}_i + \mathrm{FN}_i))} \sqrt{\sum_i((\mathrm{TN}_i + \mathrm{FP}_i)(\mathrm{TP}_i + \mathrm{FN}_i))}} = \frac{(\sum_i \pi_{ii} - \sum_i \pi_{i+} \pi_{+i})}{\sqrt{1 - \sum_i \pi_{+i}^2} \sqrt{1 - \sum_i \pi_{i+}^2}}.
	\end{eqnarray}
	The measure \(\mathrm{miM^*}\) is consistent with what was proposed as a correlation coefficient for multi-class data in Gorodkin\cite{goro}.

	\section{Variance Estimate and Confidence Interval of Single MCC}
	\subsection{Applying the delta method}
	
	In this section, we introduce two approaches for constructing asymptotic confidence intervals for a single MCC. The measures $\mathrm{maM}$, $\mathrm{miM}$, and $\mathrm{miM}^*$ are expressed as functions as follows:
	\[
	\mathrm{maM}= \varphi_1( \bm{\pi}^{(2)} ), \quad \mathrm{miM}= \varphi_2( \bm{\pi}^{(2)} ), \quad \mathrm{miM}^*= \varphi_3( \bm{\pi}^{(2)} ),
	\]
	where
	\[
	\bm{\pi}^{(2)} = (\pi_{11}, \ldots, \pi_{1r}, \pi_{21}, \ldots, \pi_{2r}, \ldots, \pi_{r1}, \ldots, \pi_{rr})^\mathsf{T},
	\]
	and $\mathsf{T}$ denotes the transpose operator. Furthermore, assuming the observed frequencies $n_{ij}\ (i, j = 1, \ldots, r)$ are obtained from a multinomial distribution with probabilities $\bm{\pi}^{(2)}$ and sample size $n = \sum_{i,j} n_{ij}$, the maximum likelihood estimator of $\pi_{ij}$ is given by
	\[
	\hat{\pi}_{ij} = \frac{n_{ij}}{n}.
	\]
	Thus, if we define the \( r^2 \times 1 \) vector 
	\[
	\hat{\bm{\pi}}^{(2)} = (\hat{\pi}_{11}, \ldots, \hat{\pi}_{1r}, \hat{\pi}_{21}, \ldots, \hat{\pi}_{2r}, \ldots, \hat{\pi}_{r1}, \ldots, \hat{\pi}_{rr})^\mathsf{T},
	\]
	then the empirical \(\mathrm{maM}\), \(\mathrm{miM}\), and \(\mathrm{miM}^*\) are given by
	\[
	\widehat{\mathrm{maM}}= \varphi_1( \hat{\bm{\pi}}^{(2)} ), \quad \widehat{\mathrm{miM}}= \varphi_2(  \hat{\bm{\pi}}^{(2)} ), \quad \widehat{\mathrm{miM}^*}= \varphi_3(  \hat{\bm{\pi}}^{(2)} ).
	\] 
	Using the central limit theorem, we have the following:
	
	\begin{gather*}
		\sqrt{n}(\hat{\bm{\pi}}^{(2)} - \bm{\pi}^{(2)}) \xrightarrow{d} N\left(\bm{0}_{r^2}, \phi(\bm{\pi}^{(2)})\right),    \ \ \ \ \text{as} \ n \rightarrow \infty\\
		\phi(\bm{\pi}^{(2)})=\mathrm{\text{diag}}(\bm{\pi}^{(2)}) - \bm{\pi}^{(2)}{\bm{\pi}^{(2)}}^\mathsf{T},
	\end{gather*}
	where $0_{r^2}$ is an $r^2$-dimensional vector with all elements equal to 0, and $\mathrm{\text{diag}}(\bm{\pi}^{(2)})$ represents a diagonal matrix of size $r^2 \times r^2$ with the vector $\bm{\pi}^{(2)}$ on its diagonal. 
	
	Thus, denoting the multiclass MCCs defined by equations (1), (2), and (3) as $\varphi_l(\bm{\pi}^{(2)})$ for $l\in1,2,3$, we apply the delta method as follows:
	
	\begin{align*}
		\sqrt{n}(\varphi_l( \hat{\bm{\pi}}^{(2)} ) - \varphi_l( \bm{\pi}^{(2)} )) &\xrightarrow{d} N \left(0, \nabla\varphi_l(  \bm{\pi}^{(2)})^\mathsf{T}\phi(\bm{\pi}^{(2)})\nabla\varphi_l(  \bm{\pi}^{(2)}) \right),    \ \ \ \ \text{as} \ n \rightarrow \infty,
	\end{align*}
	where $\nabla$ denotes the gradient operator.
	
	The derivation details of the asymptotic variances are provided in Appendix A.  For these multi-class MCCs, assuming a sufficiently large sample size $n$, the $100(1 - \alpha)\%$ confidence intervals can be approximated as follows , where $z_{\alpha/2}$ denotes the upper $\alpha/2$-quantile of the standard normal distribution:
	\[
	\varphi_l( \hat{\bm{\pi}}^{(2)} ) \pm z_{\alpha/2} \sqrt{\frac{\nabla\varphi_l( \hat{\bm{\pi}}^{(2)})^\mathsf{T}\phi(\hat{\bm{\pi}}^{(2)})\nabla\varphi_l( \hat{\bm{\pi}}^{(2)})}{n}}.
	\]

	\subsection{Fisher's z transformation}
	The core concept of MCC is Pearson's correlation coefficient. By employing Fisher's z transformation within the framework of Pearson's correlation coefficient, it is possible to account for the asymmetry in the distribution\cite{fisher}. The sampling distribution of MCC is often highly skewed. Therefore, applying Fisher's z transformation is expected to facilitate rapid convergence to a normal distribution. Define the function $f:(-1,1)\to \mathbb{R}$ as
	\[
	f(x)=\frac{1}{2}\log\left(\frac{1+x}{1-x}\right).
	\]
	Apply the delta method to the transformed equation:
	\begin{align*}
		\sqrt{n}\{f \circ \varphi_l( \hat{\bm{\pi}}^{(2)} ) - f \circ \varphi_l( \bm{\pi}^{(2)} )\} &\xrightarrow{d} N \left(0, \{\nabla f \circ \varphi_l(  \bm{\pi}^{(2)})\}^\mathsf{T}\phi(\bm{\pi}^{(2)})\{\nabla f \circ\varphi_l(  \bm{\pi}^{(2)})\} \right).
	\end{align*}
	Therefore, the $100(1-\alpha)\%$ confidence interval of a multi-class MCC can be expressed in the following form by performing the inverse transformation:
	
	\[
	f^{-1}\left( f \circ \varphi_l( \hat{\bm{\pi}}^{(2)} ) \pm z_{\alpha/2} \sqrt{\frac{\{\nabla f \circ \varphi_l(  \hat{\bm{\pi}}^{(2)})\}^\mathsf{T}\phi(\hat{\bm{\pi}}^{(2)})\{\nabla f \circ\varphi_l(  \hat{\bm{\pi}}^{(2)})\}}{n}}\right).
	\]

	\section{Variance Estimate and Confidence Interval of Paired Design}
	When evaluating the diagnostic performance of a new screening test, it is common practice to assess the same subjects using both an existing method and the desired new screening method. Similarly, in machine learning, evaluating a newly developed classification method often entails applying both existing and novel approaches to the test data. The paired design approach, where both new and existing methods are applied to the same dataset, is widely employed across diverse fields.
	
	The application of both existing and novel methods to the same data for performance evaluation holds significant importance in various domains. Statistically verifying differences between MCCs in paired designs is particularly crucial. This approach is used to compare the MCCs of two distinct methods applied to the same dataset, focusing on scenarios involving multi-class testing or classification.
	
	This section will emphasize the importance of statistically evaluating differences between MCCs in paired designs. This is critical for both academic research and practical applications in order to evaluate the performance of new screening tests, classifiers, and diagnostic tools relative to established methods.
	\begin{table}[hbtp]
		\centering
		\caption{For paired design}
		\begin{tabular}{lcccccccccccc}
			&& \multicolumn{4}{c}{$\text{True Classification:}Y= 1$} & && \multicolumn{4}{c}{$\text{True Classification:}Y=r$} \\
			\cmidrule{3-6} \cmidrule{9-12}
			&& \multicolumn{4}{c}{$\text{Prediction method 2:}X_2$} &&& \multicolumn{4}{c}{$\text{Prediction method 2:}X_2$} \\
			\cmidrule{3-6} \cmidrule{9-12}
			&& 1 & 2 & $\cdots$ & $r$ &&& 1 & 2 & $\cdots$ & $r$ \\
			\cmidrule{1-6} \cmidrule{8-12}
			$\text{Prediction method 1:} X_1$&1 &  $\pi_{111}$ & $\pi_{121}$ & $\cdots$ & $\pi_{1r1}$ & & 1 & $\pi_{11r}$ & $\pi_{12r}$ & $\cdots$ & $\pi_{1rr}$ \\
			&2 &  $\pi_{211}$ & $\pi_{221}$ & $\cdots$ & $\pi_{2r1}$ & & 2 & $\pi_{21r}$ & $\pi_{22r}$ & $\cdots$ & $\pi_{2rr}$ \\
			&$\vdots$ &  $\vdots$ & $\vdots$ & $\ddots$ & $\vdots$ &$\cdots$   & $\vdots$ & $\vdots$ & $\vdots$ & $\ddots$ & $\vdots$ \\
			&$r$ & $\pi_{r11}$ & $\pi_{r21}$ & $\cdots$ & $\pi_{rr1}$ & & $r$ & $\pi_{r1r}$ & $\pi_{r2r}$ & $\cdots$ & $\pi_{rrr}$ \\
			\cmidrule{1-6} \cmidrule{8-12}
		\end{tabular}
	\end{table}
	
	\subsection{Applying delta method}
	Consider an $r \times  r \times r$ table as depicted in Table 2 representing a general table of cell probabilities, denoted as $\pi_{ijk}$, where $i$ represents the class of prediction method 1, $j$ represents the class of prediction method 2, and $k$ represents the true classification. Assuming a random sample from the population distribution of interest, and denoting the results of prediction methods 1 and 2 as respectively $X_1$ and $X_2$, and the true classification as $Y$, the probabilities $\pi_{ijk}$ are defined by $\pi_{ijk} = P(X_1=i, X_2=j, Y=k)$, for $i,j,k = {1,2,\ldots,r}$. Using the notation $\sum_{i=1}^{r} \pi_{ijk}=\pi_{+jk}$, $\sum_{j=1}^{r} \pi_{ijk}=\pi_{i+k}$, and $\sum_{k=1}^{r} \pi_{ijk}=\pi_{ij+}$ , the true positive rate ($\mathrm{TP}_{a}$), false positive rate ($\mathrm{FP}_{a}$), false negative rate ($\mathrm{FN}_{a}$), and true negative rate ($\mathrm{TN}_{a}$) for each class $a \ (a = 1,\ldots, r)$ for prediction method 1 are defined as follows:
	\begin{eqnarray*}
		\text{True Positive (TP):} \mathrm{TP}_{1a} &=& P(X_1=a, Y=a)=\pi_{a+a},\\
		\text{False Positive (FP):} \mathrm{FP}_{1a} &=& P(X_1=a, Y\neq a)=\sum_{\substack{k=1 \\ k\neq a}}^r \pi_{a+k},\\
		\text{False Negative (FN):} \mathrm{FN}_{1a} &=& P(X_1\neq a, Y=a)=\sum_{\substack{i=1 \\ i\neq a}}^r \pi_{i+a},\\
		\text{True Negative (TN):} \mathrm{TN}_{1a} &=& P(X_1\neq a, Y\neq a)=\sum_{\substack{i=1 \\ i\neq a}}^r\sum_{\substack{k=1 \\ k\neq a}}^r \pi_{i+k}.
	\end{eqnarray*}
	Similarly, $\mathrm{TP}_a$, $\mathrm{FP}_a$, $\mathrm{FN}_a$, and $\mathrm{TN}_a$ for each class $a \ (a = 1,\ldots, r)$ for prediction method 2 are defined as follows:
	\begin{eqnarray*}
		\text{True Positive (TP):} \mathrm{TP}_{2a} &=& P(X_2=a, Y=a)=\pi_{+aa},\\
		\text{False Positive (FP):} \mathrm{FP}_{2a} &=& P(X_2=a, Y\neq a)=\sum_{\substack{k=1 \\ k\neq a}}^r \pi_{+ak},\\
		\text{False Negative (FN):} \mathrm{FN}_{2a} &=& P(X_2\neq i, Y=a)=\sum_{\substack{i=1 \\ i\neq a}}^r \pi_{+ja},\\
		\text{True Negative (TN):} \mathrm{TN}_{2a} &=& P(X_2\neq a, Y\neq a)=\sum_{\substack{j=1 \\ j\neq a}}^r\sum_{\substack{k=1 \\ k\neq a}}^r \pi_{+jk}.
	\end{eqnarray*}
	
	We explore the observed cell frequencies \(n_{ijk}\) obtained from a multinomial distribution with probabilities
	\[
	\bm{\pi}^{(3)} = (\pi_{111}, \ldots, \pi_{1r1}, \ldots, \pi_{rr1}, \ldots, \pi_{rrr})^\mathsf{T}
	\]
	and a total sample size of \(n = \sum_{i, j, k} n_{ijk}\). The maximum likelihood estimators for \(\pi_{ijk}\) are given by \(\hat{\pi}_{ijk} = n_{ijk}/n\). Thus, if we define the \( r^3 \times 1 \) vector
	\[
	\hat{\bm{\pi}}^{(3)} = (\hat{\pi}_{111}, \ldots, \hat{\pi}_{1r1}, \ldots, \hat{\pi}_{rr1}, \ldots, \hat{\pi}_{rrr})^\mathsf{T},
	\]
	we obtain the following:
	
	\begin{gather*}
		\sqrt{n}(\hat{\bm{\pi}}^{(3)} - \bm{\pi}^{(3)}) \xrightarrow{d} N\left(\bm{0}_{r^3}, \phi(\bm{\pi}^{(3)})\right),\\
		\phi(\bm{\pi}^{(3)}) = \text{diag}(\bm{\pi}^{(3)}) - \bm{\pi}^{(3)}{\bm{\pi}^{(3)}}^\mathsf{T},
	\end{gather*}
	where $0_{r^3}$ is an $r^3$-dimensional vector with all elements equal to 0, and $\mathrm{\text{diag}}(\bm{\pi^{(3)}})$ represents a diagonal matrix of size $r^3 \times r^3$ with the vector $\bm{\pi}^{(3)}$ on its diagonal. 
	
	Representing the multi-class MCCs for prediction method 1 as \(\mathrm{maM}_1\), \(\mathrm{miM}_1\), and \(\mathrm{miM}_1^*\), and those for prediction method 2 as \(\mathrm{maM}_2\), \(\mathrm{miM}_2\), and \(\mathrm{miM}_2^*\), we define the vectors \(\mathbf{M}_1 = (\mathrm{maM}_1, \mathrm{maM}_2)^\mathsf{T}\), \(\mathbf{M}_2 = (\mathrm{miM}_1, \mathrm{miM}_2)^\mathsf{T}\), and \(\mathbf{M}_3 = (\mathrm{miM}_1^*, \mathrm{miM}_2^*)^\mathsf{T}\). Expressing the three metrics  using \( l = 1, 2, 3 \), the empirical \(\mathbf{M}_l\) are denoted as \(\hat{\mathbf{M}}_l\). Based on the previous discussion, we apply distributional convergence and the delta method to obtain the following simple method:
	\[
	\sqrt{n}(\hat{\mathbf{M}}_l - \mathbf{M}_l) \xrightarrow{d} N\left(0,\nabla\mathbf{M}_l^\mathsf{T} \bm{\phi}(\bm{\pi}^{(3)}) \nabla\mathbf{M}_l \right).
	\]
	This allows for the derivation of asymptotic confidence intervals for \(\mathrm{maM}_1 - \mathrm{maM}_2\), \(\mathrm{miM}_1 - \mathrm{miM}_2\), and \(\mathrm{miM}_1^* - \mathrm{miM}_2^*\) as detailed in Appendix B.

	\subsection{Modified transformation method}
	We consider applying Fisher's z transformation in paired designs, similar to single cases , followed by the application of the delta method. However, it is known that Fisher's z transformation cannot be inverted when considering differences\cite{zou}. In the study by  Itaya et al. \cite{ita}, the following transformation was considered for biM, which enabled an explanation of the asymmetry in the distribution. Define the function $g:(-2,2)\to \mathbb{R}$ as
	\begin{gather*}
		g(x)=\frac{1}{2}\log\left(\frac{2+x}{2-x}\right).
	\end{gather*}
	In our study, by adopting the function $g(x)$, we expect to address the skewness in the sampling distribution and achieve faster convergence to a normal distribution. We define the functions \(\psi_{1}(\bm{\pi}^{(3)}) = \mathrm{maM}_1 - \mathrm{maM}_2\), \(\psi_{2}(\bm{\pi}^{(3)}) = \mathrm{miM}_1 - \mathrm{miM}_2\), and \(\psi_{3}(\bm{\pi}^{(3)}) = \mathrm{miM}^*_1 - \mathrm{miM}^*_2\). The estimators can thus be expressed as \(\psi_{l}(\hat{\bm{\pi}}^{(3)}),\ l=1,2,3\). Applying the delta method after transforming by function $g(x)$, we have the following:
	\begin{align*}
		\sqrt{n}\{g \circ \psi_{l}( \hat{\bm{\pi}}^{(3)} ) - g \circ \psi_{l}( \bm{\pi}^{(3)} )\} &\xrightarrow{d} N \left(0, \{\nabla g \circ \psi_l(  \bm{\pi}^{(3)})\}^\mathsf{T}\phi(\bm{\pi}^{(3)})\{\nabla g \circ\psi_l(  \bm{\pi}^{(3)})\} \right).
	\end{align*}
	
	Therefore, the $100(1-\alpha)\%$ confidence intervals for the differences in the multi-class MCCs between two methods can be expressed in the following form by performing the inverse transformation:
	
	\begin{align*}
		&g^{-1}\left(g \circ \psi_{l}( \hat{\bm{\pi}}^{(3)} )  \pm z_{\alpha/2} \sqrt{\frac{\{\nabla g \circ \psi_l(  \hat{\bm{\pi}}^{(3)})\}^\mathsf{T}\phi(\hat{\bm{\pi}}^{(3)})\{\nabla g \circ\psi_l(  \hat{\bm{\pi}}^{(3)})\}}{n}}\right).
	\end{align*}
	
	\section{Simulation}
	In this section, we evaluate the behavior of the methods proposed in the present paper and confirm the operating characteristics of the three defined multi-class MCCs.
	\subsection{Simulation setup}
	To evaluate the performance of the proposed confidence intervals for the multi-class MCCs herein, we computed the coverage probability over 100,000 iterations of generating 95\% confidence intervals. We set up a three-class scenario and randomly sampled from a multinomial distribution based on the cell probabilities shown in Table \ref{nomal1} for each scenario. Scenarios 1 and 2 assume nearly equal probabilities (1/3) for the true states of classes 1, 2, and 3. In contrast, Scenarios 3 and 4 represent situations where the data exhibit class imbalance.
	
	Next, we conducted simulations to evaluate the coverage probabilities of confidence intervals for the difference between the MCCs of the two prediction methods in the paired design. The data were generated according to a multinomial distribution with $r=3$ (classes 1, 2, and 3), as shown in Table \ref{paired1}. Sample sizes were set to 50, 100, 400, and 800. For each combination of true probabilities and sample sizes, 100,000 datasets were generated, and 95\% confidence intervals were calculated for each dataset. In Scenarios 1 and 2, the true probabilities for classes 1, 2, and 3 were equal (1/3). In Scenarios 3 and 4, the probabilities were biased towards specific classes. Moreover, in Scenarios 1 and 3, the MCCs of the two prediction methods were equal. On the other hand, in Scenarios 2 and 4, the performances of the two prediction methods differed.
	\begin{table}[hbtp]
		\centering
		\caption{Simulation study: True cell probabilities and MCCs}
		\begin{tabular}{cc}
			\begin{minipage}[t]{0.48\linewidth}
				\centering
				\begin{tabular}{lcccc}
					\hline 
					\rowcolor{mygray!150}Scenario 1 & & \multicolumn{3}{c}{True Classification} \\
					\rowcolor{mygray!150}	& & 1 & 2 & 3 \\
					\rowcolor{white}	& 1 & 28/100 & 2/100 & 3/100 \\
					\rowcolor{white}	Prediction & 2 & 3/100 & 28/100 & 2/100  \\
					\rowcolor{white}	& 3 & 2/100 & 3/100 & 29/100 \\
					\rowcolor{mygray!70}	\multicolumn{5}{l}{$\text{maM}=0.77, \text{miM}=0.78, \text{miM}^*=0.77$} \\
					\hline
				\end{tabular}
				\vspace{0.5cm} 
				\begin{tabular}{lcccc}
					\hline 
					\rowcolor{mygray!150}	Scenario 2 & & \multicolumn{3}{c}{True Classification} \\
					\rowcolor{mygray!150}	& & 1 & 2 & 3 \\
					\rowcolor{white}& 1 & 11/100 & 11/100 & 11/100 \\
					\rowcolor{white}	Prediction & 2 & 11/100 & 11/100 & 11/100 \\
					\rowcolor{white}	& 3 & 11/100 & 11/100 & 12/100 \\
					\rowcolor{mygray!70}	\multicolumn{5}{l}{$\text{maM}=0.01, \text{miM}=0.01, \text{miM}^*=0.01$} \\
					\hline
				\end{tabular}
			\end{minipage} 
			&
			\begin{minipage}[t]{0.48\linewidth}
				\centering
				\begin{tabular}{lcccc}
					\hline 
					\rowcolor{mygray!150}	Scenario 3 & & \multicolumn{3}{c}{True Classification} \\
					\rowcolor{mygray!150}& & 1 & 2 & 3 \\
					\rowcolor{white}	& 1 & 2/100 & 5/100 & 0/100 \\
					\rowcolor{white}	Prediction & 2 & 2/100 & 70/100 & 2/100 \\
					\rowcolor{white}	& 3 & 2/100 & 2/100 & 15/100 \\
					\rowcolor{mygray!70}	\multicolumn{5}{l}{$\text{maM}=0.59, \text{miM}=0.81, \text{miM}^*=0.67$} \\
					\hline
				\end{tabular}
				\vspace{0.5cm} 
				\begin{tabular}{lcccc}
					\hline
					\rowcolor{mygray!150}	Scenario 4 & & \multicolumn{3}{c}{True Classification} \\
					\rowcolor{mygray!150}	& & 1 & 2 & 3 \\
					\rowcolor{white}	& 1 & 2/100 & 25/100 & 6/100 \\
					\rowcolor{white}	Prediction & 2 & 11/100 & 26/100 & 6/100 \\
					\rowcolor{white}	& 3 & 2/100 & 25/100 & 6/100 \\
					\rowcolor{mygray!40}		\multicolumn{5}{l}{$\text{maM}=0.00, \text{miM}=0.01, \text{miM}^*=0.00$} \\
					\hline
				\end{tabular}
			\end{minipage}
		\end{tabular}
		\label{nomal1}
	\end{table}

	\begin{table}[hbtp]
		\centering
		\caption{Simulation study of paired design: True cell probabilities}
		\begin{tabular}{lcccccccccccc}
			\hline
			&& \multicolumn{3}{c}{True Classification=1} &&\multicolumn{3}{c}{True Classification=2}&& \multicolumn{3}{c}{True Classification=3} \\
			\cmidrule[\heavyrulewidth]{3-5} \cmidrule[\heavyrulewidth]{7-9} \cmidrule[\heavyrulewidth]{11-13}
			\rowcolor{mygray!150}Scenario 1&& \multicolumn{3}{c}{Prediction method 2} && \multicolumn{3}{c}{Prediction method 2} &&\multicolumn{3}{c}{Prediction method 2}\\
			\rowcolor{mygray!150}	&& 1 & 2 & 3 && 1 & 2 & 3 && 1 & 2 & 3 \\
			Prediction method 1&1 & 40/300 & 10/300 & 10/300 && 5/300 & 10/300& 5/300 && 5/300 & 5/300 & 10/300\\
			&2 &  10/300 & 5/300 & 5/300 && 10/300& 40/300 & 10/300 && 5/300 & 5/300 & 10/300 \\
			&3 &10/300 & 5/300 & 5/300 && 5/300 & 10/300 & 5/300 && 10/300 & 10/300 & 40/300 \\
			\rowcolor{mygray!70}\multicolumn{13}{l}{$\text{maM}_1=\text{miM}_1={\text{miM}_1}^*=0.40$} \\
			\multicolumn{13}{l}{$\text{maM}_2=\text{miM}_2={\text{miM}_2}^*=0.40$} \\
			\hline\hline
			&& \multicolumn{3}{c}{True Classification=1} &&\multicolumn{3}{c}{True Classification=2}&& \multicolumn{3}{c}{True Classification=3} \\
			\cmidrule[\heavyrulewidth]{3-5} \cmidrule[\heavyrulewidth]{7-9} \cmidrule[\heavyrulewidth]{11-13}
			\rowcolor{mygray!150}Scenario 2&& \multicolumn{3}{c}{Prediction method 2} && \multicolumn{3}{c}{Prediction method 2} &&\multicolumn{3}{c}{Prediction method 2}\\
			\rowcolor{mygray!150}	&& 1 & 2 & 3 && 1 & 2 & 3 && 1 & 2 & 3 \\
			Prediction method 1&1 & 30/300 & 15/300 & 15/300 && 5/300 & 10/300& 5/300 && 5/300 & 5/300 & 10/300\\
			&2 &  10/300 & 5/300 & 5/300 && 15/300& 30/300 & 15/300 && 5/300 & 5/300 & 10/300 \\
			&3 &10/300 & 5/300 & 5/300 && 5/300 & 10/300 & 5/300 && 15/300 & 15/300 & 30/300 \\
			\rowcolor{mygray!70}\multicolumn{13}{l}{$\text{maM}_1=\text{miM}_1={\text{miM}_1}^*=0.40$} \\
			\multicolumn{13}{l}{$\text{maM}_2=\text{miM}_2={\text{miM}_2}^*=0.25$} \\
			\hline
			\hline
			&& \multicolumn{3}{c}{True Classification=1} &&\multicolumn{3}{c}{True Classification=2}&& \multicolumn{3}{c}{True Classification=3} \\
			\cmidrule[\heavyrulewidth]{3-5} \cmidrule[\heavyrulewidth]{7-9} \cmidrule[\heavyrulewidth]{11-13}
			\rowcolor{mygray!150}Scenario 3&& \multicolumn{3}{c}{Prediction method 2} && \multicolumn{3}{c}{Prediction method 2} &&\multicolumn{3}{c}{Prediction method 2}\\
			\rowcolor{mygray!150}	&& 1 & 2 & 3 && 1 & 2 & 3 && 1 & 2 & 3 \\
			Prediction method 1&1 & 120/500 & 30/500 & 30/500 && 5/500 & 10/500 & 5/500 && 5/500 & 5/500 & 10/500\\
			&2 &  30/500 & 15/500 & 15/500 && 10/500& 40/500 & 10/500 && 5/500 & 5/500 & 10/500 \\
			&3 &30/500 & 15/500 & 15/500 && 5/500 & 10/500 & 5/500 && 10/500 & 10/500 & 40/500\\
			\rowcolor{mygray!70}	\multicolumn{13}{l}{$\text{maM}_1=0.37,\text{miM}_1=0.40,{\text{miM}_1}^*=0.37$} \\
			\multicolumn{13}{l}{$\text{maM}_2=0.37,\text{miM}_2=0.40,{\text{miM}_2}^*=0.37$} \\
			\hline
			\hline
			&& \multicolumn{3}{c}{True Classification=1} &&\multicolumn{3}{c}{True Classification=2}&& \multicolumn{3}{c}{True Classification=3} \\
			\cmidrule[\heavyrulewidth]{3-5} \cmidrule[\heavyrulewidth]{7-9} \cmidrule[\heavyrulewidth]{11-13}
			\rowcolor{mygray!150}Scenario 4&& \multicolumn{3}{c}{Prediction method 2} && \multicolumn{3}{c}{Prediction method 2} &&\multicolumn{3}{c}{Prediction method 2}\\
			\rowcolor{mygray!150}	&& 1 & 2 & 3 && 1 & 2 & 3 && 1 & 2 & 3 \\
			Prediction method 1&1 & 190/500 & 80/500 & 90/500 && 5/500 & 5/500 & 0/500 && 5/500 & 5/500 & 5/500\\
			&2 &  5/500 & 5/500 & 5/500 && 5/500& 10/500 & 5/500 && 5/500 & 5/500 & 15/500 \\
			&3 &0/500 & 5/500 & 5/500 && 5/500 & 5/500 & 5/500 && 5/500 & 5/500 & 20/500\\
			\rowcolor{mygray!70}		\multicolumn{13}{l}{$\text{maM}_1=0.48,\text{miM}_1=0.73,{\text{miM}_1}^*=0.53$} \\
			\multicolumn{13}{l}{$\text{maM}_2=0.20,\text{miM}_2=0.27,{\text{miM}_2}^*=0.20$} \\
			\hline
		\end{tabular}\\
		\label{paired1}
	\end{table}

	\subsection{Simulation results}
	Table \ref{nomal2} presents the coverage probabilities of the proposed $95\%$ confidence intervals for a single MCC for each scenario. The coverage probabilities approach $0.95$ as the sample size increases, as indicated by the $\rm{maM}$, $\rm{miM}$, and $\rm{miM}^*$ values. However, in imbalanced scenarios such as Scenarios 3 and 4, the coverage probabilities decrease in situations with small sample sizes. In particular, $\rm{maM}$ deviates considerably from the nominal value. This deviation is likely due to the macro-averaged approach treating each class equally. Additionally, both $\rm{miM}$ and $\text{miM}^*$ exhibit decreased coverage probabilities in imbalanced scenarios compared to balanced situations. However, applying Fisher's z transformation can improve the coverage probabilities. In particular, $\rm{miM}^*$ demonstrates coverage probabilities closer to the nominal value even with small sample sizes when Fisher's z transformation is applied. Among the three metrics , $\rm{miM}$ tended to have larger values than $\rm{maM}$ and $\rm{miM}^*$, especially in imbalanced situations.
	
	Table \ref{paired2} lists the coverage probabilities of the $95\%$ confidence intervals for the difference in MCCs between two prediction methods in a paired design. Similar to the confidence intervals for single MCCs , the coverage probabilities approach the nominal value as the sample size increases.
	
	\begin{table}[hbtp]
		\centering
		\caption{Estimated coverage probabilities for confidence intervals of single MCCs}
		\vspace{1mm}
		\renewcommand{\arraystretch}{1.3} 
		\begin{tabular}{cccccccc}
			\toprule
			Scenario &n& $\varphi_1(  \bm{\pi}^{(2)})$& $f \circ \varphi_1(  \bm{\pi}^{(2)})$ &  $\varphi_2(  \bm{\pi}^{(2)})$& $f \circ \varphi_2(  \bm{\pi}^{(2)})$ & $\varphi_3(  \bm{\pi}^{(2)})$& $f \circ \varphi_3(  \bm{\pi}^{(2)})$   \\
			\midrule
			1& 50 &0.9230  &0.9541  &0.9412  &0.9563  &0.9282 &0.9565 \\
			\rowcolor{mygray!70}  &  100 &0.9326  &0.9505  &0.9329  &0.9482  &0.9323 &0.9510  \\
			& 400 &0.9449  &0.9491  &0.9450  &0.9484  &0.9448 &0.9492 \\
			\rowcolor{mygray!70}  &800  &0.9496  &0.9486  &0.9510  &0.9467  &0.9499 &0.9482  \\
			\hline
			2&50 &0.9315  &0.9385  &0.9258  &0.9400  &0.9349 &0.9418 \\
			\rowcolor{mygray!70}   &100 &0.9418  &0.9444  &0.9424  &0.9424  &0.9431 &0.9457 \\
			&400 & 0.9471 &0.9482  &0.9478  &0.9478  &0.9474 &0.9483 \\
			\rowcolor{mygray!70}  &800 & 0.9489 &0.9492  &0.9480  &0.9480  &0.9489 &0.9493  \\
			\hline
			3&50 &0.8349  &0.8501  &0.8943  &0.9756  &0.9143 &0.9528 \\
			\rowcolor{mygray!70}   &100 &0.8723  &0.8820  &0.9390 &0.9481  &0.9319 &0.9505 \\
			&400 &0.9396  &0.9416  &0.9426  &0.9560  &0.9460 &0.9506 \\
			\rowcolor{mygray!70}  &800  &0.9445  &0.9450  &0.9523  &0.9485  & 0.9475&0.9495  \\
			\hline
			4&50 &0.9055  & 0.9109  &0.9247  &0.9391  &0.9228 &0.9283 \\
			\rowcolor{mygray!70}&100 &0.9287  &0.9315 &0.9411  &0.9411   &0.9382 &0.9408 \\
			&400 &0.9455   &0.9463  &0.9474  &0.9474  &0.9472 &0.9479 \\
			\rowcolor{mygray!70}&800  &0.9481  &0.9483  &0.9469  & 0.9469 &0.9485 &0.9489  \\
			\bottomrule
		\end{tabular}\\ 
		\label{nomal2}
	\end{table}

	\begin{table}[hbtp]
		\centering
		\caption{Estimated coverage probabilities for confidence intervals in paired designs}
		\vspace{1mm}
		\renewcommand{\arraystretch}{1.3} 
		\begin{tabular}{cccccccc}
			\toprule
			Scenario &n& $\psi_1(  \bm{\pi}^{(3)})$& $g \circ\psi_1(  \bm{\pi}^{(3)})$ &  $\psi_2(  \bm{\pi}^{(3)})$& $g \circ\psi_2(  \bm{\pi}^{(3)})$ & $\psi_3(  \bm{\pi}^{(3)})$&$g \circ\psi_3(  \bm{\pi}^{(3)})$   \\
			\midrule
			1& 50 & 0.9360 & 0.9387 & 0.9462  & 0.9467 & 0.9381& 0.9409 \\
			\rowcolor{mygray!70}  &  100 & 0.9431 & 0.9443 & 0.9458 & 0.9458&0.9442&0.9455  \\
			& 400 &0.9478 & 0.9481 & 0.9486 & 0.9487 & 0.9482& 0.9484 \\
			\rowcolor{mygray!70}  &800  &0.9488 &0.9489 &0.9491 & 0.9493 & 0.9487&0.9489  \\
			\hline
			2& 50 & 0.9361 & 0.9393&0.9440  & 0.9443 & 0.9386& 0.9417 \\
			\rowcolor{mygray!70}  &  100 &0.9440& 0.9454& 0.9458& 0.9484&0.9447&0.9463 \\
			& 400 &0.9486 & 0.9490 &0.9492 & 0.9498& 0.9489& 0.9494 \\
			\rowcolor{mygray!70}  &800  &0.9490 &0.9493 &0.9493 &0.9496 & 0.9490&0.9494  \\
			\hline
			3& 50 & 0.9301& 0.9331&0.9458  & 0.9463 & 0.9359& 0.9387\\
			\rowcolor{mygray!70}  &  100 &0.9420&0.9432 & 0.9478& 0.9478&0.9446&0.9460\\
			& 400 &0.9490 & 0.9494&0.9492 & 0.9493& 0.9488& 0.9491 \\
			\rowcolor{mygray!70}  &800  &0.9489 &0.9490&0.9498 &0.9500& 0.9491&0.9493 \\
			\hline
			4& 50 & 0.8974& 0.9010&0.9393  & 0.9443 & 0.9211& 0.9249\\
			\rowcolor{mygray!70}  &  100 &0.9275&0.9293& 0.9451& 0.9461&0.9360&0.9375\\
			& 400 &0.9465&0.9467 &0.9499 &0.9505& 0.9476& 0.9483 \\
			\rowcolor{mygray!70}  &800  &0.9484 &0.9486&0.9489 &0.9490& 0.9491&0.9491 \\
			\bottomrule
		\end{tabular}\\ 
		\label{paired2}
	\end{table}
	\section{Real Data Analysis}
	In this section, we present a real data analysis applying our method to the dataset provided by Takahashi et al. \cite{taka23}, focusing on their skin cancer classification system. They conducted a comparative analysis of classification performance between a region-based convolutional neural network algorithm (FRCNN) and board-certified dermatologists (BCD) using hypothesis testing based on the $\text{F}_1$ score. As shown in Table \ref{ex}, both FRCNN and BCD classify lesions into six categories (malignant melanoma, basal cell carcinoma, nevus, seborrheic keratosis, senile lentigo , and hematoma/hemangioma). To compare the classification performance of FRCNN and BCD from the perspective of MCC, we employed a simple method to derive asymptotic confidence intervals for the differences.
	
	The results are presented in Table \ref{real}. Since the lower confidence limits of the differences in $\text{maM}$, $\text{miM}$, and $\text{miM}^*$ are all greater than 0, we can conclude that there are significant differences at the 5\% significance level in terms of statistical hypothesis testing. Our results are consistent with the results based on the $\text{F}_1$ scores shown in Takahashi et al.\cite{taka23}. Specifically, the classification of six classes by FRCNN was significantly superior to that by BCD.

	\begin{table}[hbtp]
		\centering
		\caption{Example data}
		\vspace{1mm}
		\renewcommand{\arraystretch}{1.3} 
		\begin{tabular}{ccccccc}
			\toprule
			& \multicolumn{6}{c}{True Classification} \\
			\rowcolor{mygray!150} FRCN & MM & BCC & Nevus & SK & H/H & SL \\
			\hline
			MM & 327 & 6 & 42 & 21 & 3 & 0 \\
			\rowcolor{mygray!70} BCC & 9 & 108 & 6 & 9 & 0 & 0 \\
			Nevus & 48 & 12 & 967 & 36 & 18 & 0 \\
			\rowcolor{mygray!70} SK & 21 & 6 & 20 & 223 & 0 & 3 \\
			H/H & 0 & 0 & 3 & 0 & 57 & 0 \\
			\rowcolor{mygray!70} SL & 3 & 0 & 0 & 0 & 0 & 42 \\
			\bottomrule
			& \multicolumn{6}{c}{True Classification} \\
			\rowcolor{mygray!150} BCD & MM & BCC & Nevus & SK & H/H & SL \\
			\hline
			MM & 340 & 10 & 131 & 18 & 9 & 0 \\
			\rowcolor{mygray!70} BCC & 12 & 104 & 11 & 24 & 1 & 1 \\
			Nevus & 22 & 3 & 823 & 17 & 6 & 0 \\
			\rowcolor{mygray!70} SK & 26 & 14 & 68 & 225 & 1 & 7 \\
			H/H & 3 & 1 & 11 & 0 & 61 & 0 \\
			\rowcolor{mygray!70} SL & 5 & 0 & 4 & 5 & 0 & 37 \\
			\bottomrule
		\end{tabular}
		\label{ex}
	\end{table}

	\begin{table}[hbtp]
		\centering
		\caption{Real data analysis}
		\vspace{1mm}
		\renewcommand{\arraystretch}{1.3} 
		\begin{tabular}{ccccc}
			\toprule
			&FRCN & BCD & Difference & 95\%CI (Lower, Upper)    \\
			\midrule
			$\text{maM}$& 0.812 & 0.723&0.089&(0.056, 0.122)  \\
			\rowcolor{mygray!70} $\text{miM}$& 0.834 & 0.754 & 0.080&(0.056, 0.105) \\
			$\text{miM}^*$ &0.788&0.708&0.079&(0.050, 0.108)  \\
			\bottomrule
		\end{tabular}\\ 
		\label{real}
	\end{table}
	
	\section{Discussion}
	Despite the widespread use of MCC as a performance metric for classification and diagnosis across various fields, discussion regarding its statistical utility has been limited. Particularly for multi-class MCC, there is no universally accepted definition established. However, despite the lack of a clear definition , some studies have employed evaluations based on macro-averaged or micro-averaged metrics such as $\text{F}_1$ scores to assess the performance of multi-class classification\cite{tal,li,sam,ahu,tra,alle}. Additionally, the $R_k$ statistic has been proposed as an extension of MCC to the multi-class setting by Gorodkin\cite{goro}, but none of these studies have taken up the subject of statistical inference . 
	
	In the present study, we organized the  macro-averaged and micro-averaged metrics, which are commonly used by practitioners to evaluate the performance with MCC  in multi-class situations. The macro-averaged metric was defined as the average of the class MCCs , treating all classes equally. The micro-averaged MCC was defined in two ways: by aggregating the contributions of all classes and calculating the mean metric ($\text{miM}$), and by taking the mean of the variance and covariance based on Pearson's correlation coefficient ($\text{miM}^*$). miM is composed solely of the diagonal elements and has a proportional relationship with accuracy, and so does not preserve the advantages of MCC. Therefore, it is important to note that in situations where other metrics such as accuracy and the $\text{F}_1$ score may provide overly optimistic evaluations, miM might also similarly overestimate .
	We derived the asymptotic variances for the three defined metrics by applying the delta method. Additionally, we explored an approach involving Fisher's z transformation to account for the asymmetry in the distribution of MCC.
	
	Furthermore, aiming to compare two MCCs, we proposed a method for deriving confidence intervals for differences based on a paired design. Given the property that Fisher's z transformation cannot be inverted for differences, we proposed an approach of applying  modified transformation to the differences of MCCs. 
	
	Simulation results consistently showed that the coverage probabilities of all three types of multi-class MCCs approached nominal values as sample size increased. Similar results were obtained for Fisher's z transformation, making it inconclusive which method is superior. 
	
	One limitation of this study is that the proposed procedures rely on large-sample theory and require a large sample size to appropriately work. For two-sample designs, the MCCs can be compared by setting the covariance part to zero.

	Overall, further research is warranted to explore alternative approaches and address the limitations of our study. In the present study, nominal classification was assumed, but in actual practice, there may be situations in which the classification has an order. When evaluating the performance of classification for ordinal categories using a multi-class MCC, statistical inference based on an index that takes the ordinal categories into account may be able to increase the detection power. Therefore, it is also necessary to develop a metric to evaluate the classification performance for ordinal categories.
	
	\section{Acknowledgement} This work was supported by JST SPRING, Japan Grant Number JPMJSP2179.
	\section{Supplemental Material}
	The R code to reproduce the simulation is available on GitHub.\\
	\textbf{GitHub link}: https://github.com/tamuraj

	\newpage	
	\appendix
	\section*{Appendix A: Derivation of Asymptotic Variance}
	\subsection*{Asymptotic variance of $\widehat{\mathrm{maM}}$}
	The derivation of the asymptotic variance of $\widehat{\mathrm{maM}}$ is, using the multivariate delta method for $\bm{\pi}^{(2)}$, as follows:
	\begin{align*}
		\mathrm{maM}= \varphi_1( \bm{\pi}^{(2)} )&,\ \widehat{\mathrm{maM}}=\varphi_1( \hat{\bm{\pi}}^{(2)} ),\\
		\sqrt{n}(\varphi_1( \hat{\bm{\pi}}^{(2)} ) - \varphi_1( \bm{\pi}^{(2)} )) &\xrightarrow{d} N \left(0, \nabla\varphi_1(  \bm{\pi}^{(2)})^\mathsf{T}\phi(\bm{\pi}^{(2)})\nabla\varphi_1(  \bm{\pi}^{(2)}) \right),\\
		\nabla\varphi_1(  \bm{\pi}^{(2)})&=\left[\frac{\partial (\varphi_1(  \bm{\pi}^{(2)}))}{\partial (\bm{\pi}^{(2)})}\right].
	\end{align*}
	We have 	\begin{align*}
		A_{ii}^{\mathrm{maM}} &= \frac{\partial (\varphi_1(  \bm{\pi}^{(2)}))}{\partial (\pi_{ii})} \\
		&= \frac{1}{r} \left\{
		\frac{1-\pi_{i+}-\pi_{+i}}{\sqrt{\pi_{i+}\pi_{+i}(1-\pi_{i+})(1-\pi_{+i})}}  -\frac{(\pi_{ii}-\pi_{+i}\pi_{i+})\left[ \left\{  (\pi_{+i}+\pi_{i+})(1-\pi_{+i})-\pi_{i+}\pi_{+i} \right\}(1-\pi_{i+})-\pi_{i+}\pi_{+i}(1-\pi_{+1})  \right]
		}{2\{\pi_{+i}\pi_{i+}(1-\pi_{+i})(1-\pi_{i+})\}^{3/2}} \right\},\\
		A_{ij}^{\mathrm{maM}} &= \frac{\partial (\varphi_1(  \bm{\pi}^{(2)}))}{\partial (\pi_{ij})} \\
		&= \frac{1}{r} \left\{ \frac{-\pi_{j+}}{\sqrt{\pi_{j+}\pi_{+j}(1-\pi_{j+})(1-\pi_{+j})}}  - \frac{(\pi_{jj}-\pi_{j+}\pi_{+j})\{\pi_{j+}(1-\pi_{j+})(1-\pi_{+j})-\pi_{j+}\pi_{+j}(1-\pi_{j+})\}}
		{2\{\pi_{j+}\pi_{+j}(1-\pi_{j+})(1-\pi_{+j})\}^{3/2}} \right. \\
		&\quad +\left. \frac{-\pi_{+i}}{\sqrt{\pi_{i+}\pi_{+i}(1-\pi_{i+})(1-\pi_{+i})}}  - \frac{(\pi_{ii}-\pi_{i+}\pi_{+i})\{\pi_{+i}(1-\pi_{i+})(1-\pi_{+i})-\pi_{i+}\pi_{+i}(1-\pi_{+i})\}}
		{2\{\pi_{+i}\pi_{i+}(1-\pi_{+i})(1-\pi_{i+})\}^{3/2}} \right\},\\
		&i,j=1,\ldots,r;i\neq j,\\
		\bm{A}_{\mathrm{maM}}&=\frac{\partial(\varphi_1(  \bm{\pi}^{(2)}))}{\partial (\bm{\pi}^{(2)})}=(A_{11}^{\mathrm{maM}}, A_{12}^{\mathrm{maM}},\ldots, A_{rr}^{\mathrm{maM}})^\mathsf{T},
	\end{align*}

	and
	\begin{align*}
		\bm{A}_{\mathrm{maM}}^\mathsf{T}(\text{diag}(\bm{\pi}^{(2)}))\bm{A}_{\mathrm{maM}}\
		&=\sum_{i=1}^{r}\sum_{j=1}^{r}\pi_{ij} (A_{ij}^{\mathrm{maM}} )^2,\\
		\bm{A}_{\mathrm{maM}}^\mathsf{T}(\bm{\pi}^{(2)}{\bm{\pi}^{(2)}}^\mathsf{T}) \bm{A}_{\mathrm{maM}} \
		&=\left\{\sum_{i=1}^{r}\sum_{j=1}^{r}\pi_{ij} (A_{ij}^{\mathrm{maM}})\right\}^2.
	\end{align*}
	Thus,
	\begin{align*}
		\nabla\varphi_1(  \bm{\pi}^{(2)})^\mathsf{T}\phi(\bm{\pi}^{(2)})\nabla\varphi_1(  \bm{\pi}^{(2)})=\sum_{i=1}^{r}\sum_{j=1}^{r}\pi_{ij} (A_{ij}^{\mathrm{maM}} )^2-\left\{\sum_{i=1}^{r}\sum_{j=1}^{r}\pi_{ij} (A_{ij}^{\mathrm{maM}} )\right\}^2.
	\end{align*}

	\subsection*{Asymptotic variance of $\widehat{\mathrm{miM}}$}
	In a similar manner to $\widehat{\mathrm{maM}}$, we can obtain using the multivariate delta-method the following: 
	\begin{align*}
		\mathrm{miM}= \varphi_2( \bm{\pi}^{(2)} )&,\ \widehat{\mathrm{miM}}=\varphi_2( \hat{\bm{\pi}}^{(2)} ),\\
		\sqrt{n}(\varphi_2( \hat{\bm{\pi}}^{(2)} ) - \varphi_2( \bm{\pi}^{(2)} )) &\xrightarrow{d} N \left(0, \nabla\varphi_2(  \bm{\pi}^{(2)})^\mathsf{T}\phi(\bm{\pi}^{(2)})\nabla\varphi_2(  \bm{\pi}^{(2)}) \right),\\
		\nabla\varphi_2(  \bm{\pi}^{(2)})&=\left[\frac{\partial (\varphi_2( \hat{\bm{\pi}}^{(2)} )}{\partial (\bm{\pi}^{(2)})}\right].
	\end{align*}
	We have 	\begin{align*}
		\frac{\partial (\varphi_2( \hat{\bm{\pi}}^{(2)} )}{\partial (\pi_{ii})}&=\frac{r}{r-1}, \quad\forall i=1,\ldots,r \;\mathrm{and}\\
		\frac{\partial (\varphi_2( \hat{\bm{\pi}}^{(2)} )}{\partial (\pi_{ij})}&=0, \quad \mathrm{if}\; i\neq j.
	\end{align*}
	Furthermore,
	\begin{align*}
		\frac{\partial (\varphi_2( \hat{\bm{\pi}}^{(2)} )}{\partial (\bm{\pi}^{(2)})}&=\left(\frac{r}{r-1},0,\ldots,0,0,\frac{r}{r-1},0.\ldots,0,\ldots,0,\frac{r}{r-1}\right)^\mathsf{T},
	\end{align*}			
	
	\begin{align*}
		\left[\frac{\partial (\varphi_2( \hat{\bm{\pi}}^{(2)} )}{\partial (\bm{\pi}^{(2)})}\right]^\mathsf{T}(\text{diag}(\bm{\pi}^{(2)}))\left[\frac{\partial (\varphi_2( \hat{\bm{\pi}}^{(2)} )}{\partial (\bm{\pi}^{(2)})}\right]\\
		&=\left(\pi_{11}\frac{r}{r-1},0,\ldots, \pi_{22}\frac{r}{r-1},0,\ldots, \pi_{rr}\frac{r}{r-1}\right)\left[\frac{\partial (\varphi_2( \hat{\bm{\pi}}^{(2)} )}{\partial (\bm{\pi}^{(2)})}\right]\\
		&=\left(\frac{r}{r-1}\right)^2\sum_{i=1}^r \pi_{ii},
	\end{align*}	
	\begin{align*}
		\left[\frac{\partial (\varphi_2( \hat{\bm{\pi}}^{(2)} )}{\partial (\bm{\pi}^{(2)})}\right]^\mathsf{T}(\bm{\pi}^{(2)}{\bm{\pi}^{(2)}}^\mathsf{T})\left[\frac{\partial (\varphi_2( \hat{\bm{\pi}}^{(2)} )}{\partial (\bm{\pi}^{(2)})}\right]\\
		&=\left(\pi_{11}\frac{r}{r-1}\sum_{i=1}^r \pi_{ii},\pi_{12}\frac{r}{r-1}\sum_{i=1}^r \pi_{ii},\ldots,  \pi_{rr}\frac{r}{r-1}\sum_{i=1}^r \pi_{ii}\right)\left[\frac{\partial (\varphi_2( \hat{\bm{\pi}}^{(2)} )}{\partial (\bm{\pi})}\right]\\
		&=\left(\frac{r}{r-1}\right)^2\left(\sum_{i=1}^r \pi_{ii}\right)^2,
	\end{align*}	
	
	\begin{align*}
		\nabla\varphi_2(  \bm{\pi}^{(2)})^\mathsf{T}\phi(\bm{\pi}^{(2)})\nabla\varphi_2(  \bm{\pi}^{(2)})
		&=\left(\frac{r}{r-1}\right)^2\left(\sum_{i=1}^r \pi_{ii}\right)\left(1-\sum_{i=1}^r \pi_{ii}\right).
	\end{align*}

	\subsection*{Asymptotic variance of $\widehat{\mathrm{miM^*}}$}
	In a similar way to the previous two metrics, we can derive the asymptotic variance of $\widehat{\mathrm{miM^*}}$: 	\begin{align*}
		\mathrm{miM}^*= \varphi_3( \bm{\pi}^{(2)} )&,\ \widehat{\mathrm{miM}}^*=\varphi_3( \hat{\bm{\pi}}^{(2)} ),\\
		\sqrt{n}(\varphi_3( \hat{\bm{\pi}}^{(2)} ) - \varphi_3( \bm{\pi}^{(2)} )) &\xrightarrow{d} N \left(0, \nabla\varphi_3(  \bm{\pi}^{(2)})^\mathsf{T}\phi(\bm{\pi}^{(2)})\nabla\varphi_3(  \bm{\pi}^{(2)}) \right),\\
		\nabla\varphi_3(  \bm{\pi}^{(2)})&=\left[\frac{\partial (\varphi_3(  \bm{\pi}^{(2)}))}{\partial (\bm{\pi}^{(2)})}\right].
	\end{align*}
	We have 	\begin{align*}
		A_{ii}^{\mathrm{miM^*}} &= \frac{\partial (\varphi_3(  \bm{\pi}^{(2)}))}{\partial (\pi_{ii})} \\
		&= \frac{1-\pi_{i+}-\pi_{+i}}{\sqrt{1-\sum_{k=1}^r{\pi_{+k}^2}}\sqrt{1-\sum_{k=1}^r{\pi_{k+}^2}}} + (\sum_{k=1}^r\pi_{kk}-\sum_{k=1}^r\pi_{k+}\pi_{+k})\left(\frac{\pi_{i+}}{\sqrt{1-\sum_{k=1}^r\pi_{+k}^2}\sqrt{(1-\sum_{k=1}^r\pi_{k+}^2)^3}}\right. \\
		&\quad + \left.\frac{\pi_{+i}}{\sqrt{1-\sum_{k=1}^r\pi_{k+}^2}\sqrt{(1-\sum_{k=1}^r\pi_{+k}^2)^3}}\right), \\
		A_{ij}^{\mathrm{miM^*}} &= \frac{\partial (\varphi_3(  \bm{\pi}^{(2)}))}{\partial (\pi_{ij})} \\
		&= \left\{\frac{-\pi_{+i}-\pi_{j+}}{\sqrt{1-\sum_{k=1}^r{\pi_{+k}^2}}\sqrt{1-\sum_{k=1}^r{\pi_{k+}^2}}} + (\sum_{k=1}^r\pi_{kk}-\sum_{k=1}^r\pi_{k+}\pi_{+k})\left(\frac{\pi_{i+}}{\sqrt{1-\sum_{k=1}^r\pi_{+k}^2}\sqrt{(1-\sum_{k=1}^r\pi_{k+}^2)^3}}\right.\right. \\
		&\quad + \left.\left.\frac{\pi_{+j}}{\sqrt{1-\sum_{k=1}^r\pi_{k+}^2}\sqrt{(1-\sum_{k=1}^r\pi_{+k}^2)^3}}\right)\right\}, \\
		&i,j=1,\ldots,r;i\neq j,\\
		\bm{A}_{\mathrm{miM^*}}&=\frac{\partial(\varphi_3(  \bm{\pi}^{(2)}))}{\partial (\bm{\pi}^{(2)})}=(A_{11}^{\mathrm{miM^*}}, A_{12}^{\mathrm{miM^*}},\ldots, A_{rr}^{\mathrm{miM^*}})^\mathsf{T}.
	\end{align*}
	Thus,
	\begin{align*}
		&\nabla\varphi_3(  \bm{\pi}^{(2)})^\mathsf{T}\phi(\bm{\pi}^{(2)})\nabla\varphi_3(  \bm{\pi}^{(2)})=\sum_{i=1}^{r}\sum_{j=1}^{r}\pi_{ij} (A_{ij}^{\mathrm{miM^*}} )^2-\left\{\sum_{i=1}^{r}\sum_{j=1}^{r}\pi_{ij} (A_{ij}^{\mathrm{miM^*}} )\right\}^2.
	\end{align*}
	\subsection*{Asymptotic variance with Fisher's z transform}
	\[
	\sqrt{n}\{f \circ \varphi_l( \hat{\bm{\pi}}^{(2)} ) - f \circ \varphi_l( \bm{\pi}^{(2)} )\} \xrightarrow{d} N \left(0, \{\nabla f \circ \varphi_l(  \bm{\pi}^{(2)})\}^\mathsf{T}\phi(\bm{\pi}^{(2)})\{\nabla f \circ\varphi_l(  \bm{\pi}^{(2)})\} \right).
	\]
	If we perform Fisher's z transformation on $\mathrm{maM},\  \mathrm{miM},\  \text{and}\  \mathrm{miM}^*$, we can derive the asymptotic variance of each to determine the asymptotic variance for the transformed states:
	\[
	\nabla f\circ \varphi_l(\bm{\pi}^{(2)}) = f'(\varphi_l(\bm{\pi}^{(2)})) \nabla \varphi_l(\bm{\pi}^{(2)})= \frac{1}{1-\varphi_l(\bm{\pi}^{(2)})^2}\nabla \varphi_l(\bm{\pi}^{(2)}), \quad l=1,2,3.
	\]
	\section*{Appendix B: Derivation of Distribution and Variance for Paired Design}
	\subsection*{Asymptotic variance of $(\widehat{\mathrm{maM}}_1-\widehat{\mathrm{maM}}_2)$}
	The derivation of the asymptotic variance of $\ (\widehat{\mathrm{maM}}_1-\widehat{\mathrm{maM}}_2)$ is as follows:
	\begin{gather*}
		\sqrt{n}(\hat{\mathbf{M}}_1- \mathbf{M}_1) \xrightarrow{d} N\left(0,\nabla\mathbf{M}_1^\mathsf{T} \bm{\phi}(\bm{\pi}^{(3)}) \nabla\mathbf{M}_1 \right),\\ 
		\bm{\mathrm{M}_1} =
		\begin{pmatrix}
			\mathrm{maM}_1 \\
			\mathrm{maM}_2
		\end{pmatrix},
		\ \nabla\mathbf{M}_1 =\frac{\partial(\bm{\mathrm{M}_1})}{\partial (\bm{\pi}^{(3)})}
	\end{gather*}
	\begin{align*}
		B_{iji}^{\mathrm{maM}_1} &= \frac{\partial (\mathrm{maM}_1)}{\partial (\pi_{iji})} \\
		&= \frac{1}{r} \left\{
		\frac{1-\pi_{i++}-\pi_{++i}}{\sqrt{\pi_{i++}\pi_{++i}(1-\pi_{i++})(1-\pi_{++i})}} -\frac{(\pi_{i+i}-\pi_{++i}\pi_{i++})\left\{((\pi_{++i}+\pi_{i++})(1-\pi_{i++})-\pi_{i++}\pi_{++i})-\pi_{i++}\pi_{++i}(1-\pi_{++i})\right\}}{2\{\pi_{++i}\pi_{i++}(1-\pi_{++i})(\pi_{i++})\}^{3/2}}\right\}, \\
		B_{ijk}^{\mathrm{maM}_1} &= \frac{\partial (\mathrm{maM}_1)}{\partial (\pi_{ijk})} \\
		&=  \frac{1}{r} \left\{\frac{-\pi_{k++}}{\sqrt{\pi_{k++}\pi_{++k}(1-\pi_{k++})(1-\pi_{++k})}}  - \frac{(\pi_{k+k}-\pi_{k++}\pi_{++k})\{\pi_{k++}(1-\pi_{k++})(1-\pi_{++k})-\pi_{k++}\pi_{++k}(1-\pi_{k++})\}}
		{2\{\pi_{k++}\pi_{++k}(1-\pi_{k++})(1-\pi_{++k})\}^{3/2}} \right. \\
		&\quad + \left. \frac{-\pi_{++i}}{\sqrt{\pi_{i++}\pi_{++i}(1-\pi_{i++})(1-\pi_{++i})}}  - \frac{(\pi_{i+i}-\pi_{i++}\pi_{++i})\{\pi_{++i}(1-\pi_{i++})(1-\pi_{++i})-\pi_{i++}\pi_{++i}(1-\pi_{++i})\}}
		{2\{\pi_{++i}\pi_{i++}(1-\pi_{++i})(1-\pi_{i++})\}^{3/2}} \right\},\\
		&i,j,k=1,\ldots,r;i\neq k, \\
		\bm{B}_{\mathrm{maM}_1}&=\frac{\partial(\mathrm{maM}_1)}{\partial (\bm{\pi}^{(3)})}=(B_{111}^{\mathrm{maM}_1}, B_{121}^{\mathrm{maM}_1},\ldots, B_{rrr}^{\mathrm{maM}_1})^\mathsf{T},
	\end{align*}
	and
	\begin{align*}
		B_{jii}^{\mathrm{maM}_2} &= \frac{\partial (\mathrm{maM}_2)}{\partial (\pi_{jii})} \\
		&= \frac{1}{r} \left\{
		\frac{1-\pi_{+i+}-\pi_{++i}}{\sqrt{\pi_{+i+}\pi_{++i}(1-\pi_{+i+})(1-\pi_{++i})}} -\frac{(\pi_{+ii}-\pi_{++i}\pi_{+i+})\left\{((\pi_{++i}+\pi_{+i+})(1-\pi_{+i+})-\pi_{+i+}\pi_{++i})-\pi_{+i+}\pi_{++i}(1-\pi_{++i})\right\}}{2\{\pi_{++i}\pi_{+i+}(1-\pi_{++i})(1-\pi_{+i+})\}^{3/2}}\right\}, \\
		B_{ijk}^{\mathrm{maM}_2} &= \frac{\partial (\mathrm{maM}_2)}{\partial (\pi_{ijk})} \\
		&=  \frac{1}{r} \left\{ \frac{-\pi_{+k+}}{\sqrt{\pi_{+k+}\pi_{++k}(1-\pi_{+k+})(1-\pi_{++k})}} - \frac{(\pi_{+kk}-\pi_{+k+}\pi_{++k})\{\pi_{+k+}(1-\pi_{+k+})(1-\pi_{++k})-\pi_{+k+}\pi_{++k}(1-\pi_{+k+})\}}
		{2\{\pi_{+k+}\pi_{++k}(1-\pi_{+k+})(1-\pi_{++k})\}^{3/2}} \right. \\
		&\quad + \left. \frac{-\pi_{++j}}{\sqrt{\pi_{+j+}\pi_{++j}(1-\pi_{+j+})(1-\pi_{+j+})}}  - \frac{(\pi_{+jj}-\pi_{+j+}\pi_{+j+})\{\pi_{+j+}(1-\pi_{+j+})(1-\pi_{+j+})-\pi_{+j+}\pi_{+j+}(1-\pi_{+j+})\}}
		{2\{\pi_{+j+}\pi_{+j+}(1-\pi_{+j+})(1-\pi_{+j+})\}^{3/2}} \right\},\\
		&i,j,k=1,\ldots,r;j\neq k,\\
		\bm{B}_{\mathrm{maM}_2}&=\frac{\partial(\mathrm{maM}_2)}{\partial (\bm{\pi}^{(3)})}=(B_{111}^{\mathrm{maM}_2},B_{121}^{\mathrm{maM}_2},\ldots, B_{rrr}^{\mathrm{maM}_2})^\mathsf{T}.
	\end{align*}
	
	\begin{align*}
		\left[\frac{\partial(\bm{\mathrm{M}_1})}{\partial (\bm{\pi}^{(3)})}\right]^\mathsf{T}\bm{\phi}(\bm{\pi}^{(3)})\left[\frac{\partial(\bm{\mathrm{M}_1})}{\partial (\bm{\pi}^{(3)})}\right]=\begin{pmatrix}
			X_{\mathrm{maM}}& Z_{\mathrm{maM}} \\
			Z_{\mathrm{maM}}& Y_{\mathrm{maM}}
		\end{pmatrix},
	\end{align*}
	where
	\begin{align*}
		X_{\mathrm{maM}} &=\sum_{i=1}^{r}\sum_{j=1}^{r}\sum_{k=1}^{r}\pi_{ijk} (B_{ijk}^{\mathrm{maM}_1} )^2-\left\{\sum_{i=1}^{r}\sum_{j=1}^{r}\sum_{k=1}^{r}\pi_{ijk} B_{ijk}^{\mathrm{maM}_1} \right\}^2,\\
		Y_{\mathrm{maM}} &=\sum_{i=1}^{r}\sum_{j=1}^{r}\sum_{k=1}^{r}\pi_{ijk} (B_{ijk}^{\mathrm{maM}_2} )^2-\left\{\sum_{i=1}^{r}\sum_{j=1}^{r}\sum_{k=1}^{r}\pi_{ijk} B_{ijk}^{\mathrm{maM}_2} \right\}^2,\\
		Z_{\mathrm{maM}} &=\sum_{i=1}^{r}\sum_{j=1}^{r}\sum_{k=1}^{r}\pi_{ijk} B_{ijk}^{\mathrm{maM}_1}B_{ijk}^{\mathrm{maM}_2} -\left\{\sum_{i=1}^{r}\sum_{j=1}^{r}\sum_{k=1}^{r}\pi_{ijk} B_{ijk}^{\mathrm{maM}_1} \sum_{i=1}^{r}\sum_{j=1}^{r}\sum_{k=1}^{r}\pi_{ijk}B_{ijk}^{\mathrm{maM}_2} \right\}.
	\end{align*}
	Therefore, the asymptotic variance of $(\widehat{\mathrm{maM}}_1-\widehat{\mathrm{maM}}_2)$ is
	\begin{align*}
		(X_{\mathrm{maM}}+Y_{\mathrm{maM}}-2Z_{\mathrm{maM}})/n.
	\end{align*}

	\subsection*{Asymptotic variance of $(\widehat{\mathrm{miM}}_1-\widehat{\mathrm{miM}}_2)$}
	The derivation of the asymptotic variance of $(\widehat{\mathrm{miM}}_1-\widehat{\mathrm{miM}}_2)$ is as follows:
	\begin{gather*}
		\sqrt{n}(\hat{\mathbf{M}}_2- \mathbf{M}_2) \xrightarrow{d} N\left(0,\nabla\mathbf{M}_2^\mathsf{T} \bm{\phi}(\bm{\pi}^{(3)}) \nabla\mathbf{M}_2\right),\\ 
		\bm{\mathrm{M}_2} =
		\begin{pmatrix}
			\mathrm{miM}_1 \\
			\mathrm{miM}_2
		\end{pmatrix},
		\ \nabla\mathbf{M}_2 =\frac{\partial(\bm{\mathrm{M}_2})}{\partial (\bm{\pi}^{(3)})}
	\end{gather*}
	\begin{align*}
		B_{iji}^{\mathrm{miM}_1} &= \frac{\partial (\mathrm{miM}_1)}{\partial (\pi_{iji})},= \frac{r}{r-1} ,\\
		B_{ijk}^{\mathrm{miM}_1} &= \frac{\partial (\mathrm{miM}_1)}{\partial (\pi_{ijk})}= 0,\\
		&i,j,k=1,\ldots,r;i\neq k,\\
		\bm{B}_{\mathrm{miM}_1}&=\frac{\partial(\mathrm{miM}_1)}{\partial (\bm{\pi}^{(3)})}=(B_{111}^{\mathrm{miM}_1}, B_{121}^{\mathrm{miM}_1},\ldots, B_{rrr}^{\mathrm{miM}_1})^\mathsf{T},
	\end{align*}
	and
	\begin{align*}
		B_{jii}^{\mathrm{miM}_2} &= \frac{\partial (\mathrm{miM}_2)}{\partial (\pi_{jii})},= \frac{r}{r-1} ,\\
		B_{ijk}^{\mathrm{miM}_2} &= \frac{\partial (\mathrm{miM}_2)}{\partial (\pi_{ijk})}= 0,\\
		&i,j,k=1,\ldots,r;j\neq k,\\
		\bm{B}_{\mathrm{miM}_2}&=\frac{\partial(\mathrm{miM}_2)}{\partial (\bm{\pi}^{(3)})}=(B_{111}^{\mathrm{miM}_2}, B_{121}^{\mathrm{miM}_2},\ldots, B_{rrr}^{\mathrm{miM}_2})^\mathsf{T}.
	\end{align*}
	
	\begin{align*}
		\left[\frac{\partial(\bm{\mathrm{M}_2})}{\partial (\bm{\pi}^{(3)})}\right]^\mathsf{T}\bm{\phi}(\bm{\pi}^{(3)})\left[\frac{\partial(\bm{\mathrm{M}_2})}{\partial (\bm{\pi}^{(3)})}\right]=\begin{pmatrix}
			X_{\mathrm{miM}}& Z_{\mathrm{miM}} \\
			Z_{\mathrm{miM}}& Y_{\mathrm{miM}}
		\end{pmatrix},
	\end{align*}
	where
	\begin{align*}
		X_{\mathrm{miM}} &= \left(\frac{r}{r-1}\right)^2 \left(\sum_{i=1}^{r} \sum_{j=1}^{r} \pi_{iji}\right)\left(1-\sum_{i=1}^{r} \sum_{j=1}^{r} \pi_{iji}\right),\\
		Y_{\mathrm{miM}} &= \left(\frac{r}{r-1}\right)^2 \left(\sum_{i=1}^{r} \sum_{j=1}^{r} \pi_{jii}\right)\left(1-\sum_{i=1}^{r} \sum_{j=1}^{r} \pi_{jii}\right),\\
		Z_{\mathrm{miM}} &= \left(\frac{r}{r-1}\right)^2\left\{\left(\sum_{i=1}^r \pi_{iii}\right)-\left(\sum_{i=1}^{r} \sum_{j=1}^{r} \pi_{iji}\right)\left(\sum_{i=1}^{r} \sum_{j=1}^{r} \pi_{jii}\right)\right\}.
	\end{align*}
	Therefore, the asymptotic variance of $(\widehat{\mathrm{miM}}_1-\widehat{\mathrm{miM}}_2)$ is
	\begin{align*}
		(X_{\mathrm{miM}}+Y_{\mathrm{miM}}-2Z_{\mathrm{miM}})/n.
	\end{align*}

	\subsection*{Asymptotic variance of $(\widehat{\mathrm{miM}^*}_1-\widehat{\mathrm{miM}^*}_2)$}
	The derivation of the asymptotic variance of $(\widehat{\mathrm{miM}^*}_1-\widehat{\mathrm{miM}^*}_2)$ is as follows:
	\begin{gather*}
		\sqrt{n}(\hat{\mathbf{M}}_3- \mathbf{M}_3) \xrightarrow{d} N\left(0,\nabla\mathbf{M}_3^\mathsf{T} \bm{\phi}(\bm{\pi}^{(3)}) \nabla\mathbf{M}_3\right),\\ 
		\bm{\mathrm{M}_3} =
		\begin{pmatrix}
			\mathrm{miM}^*_1 \\
			\mathrm{miM}^*_2
		\end{pmatrix},
		\ \nabla\mathbf{M}_3 =\frac{\partial(\bm{\mathrm{M}_3})}{\partial (\bm{\pi}^{(3)})}
	\end{gather*}
	\begin{align*}
		B_{iji}^{\mathrm{miM}^*_1} &= \frac{\partial (\mathrm{miM}^*_1)}{\partial (\pi_{iji})} \\
		&= \frac{1-\pi_{i++}-\pi_{++i}}{\sqrt{1-\sum_{l=1}^r{\pi_{++l}^2}}\sqrt{1-\sum_{l=1}^r{\pi_{l++}^2}}} + (\sum_{l=1}^r \pi_{l+l}-\sum_{l=1}^r\pi_{l++}\pi_{++l})\left(\frac{\pi_{i++}}{\sqrt{1-\sum_{l=1}^r\pi_{++l}^2}\sqrt{(1-\sum_{l=1}^r\pi_{l++}^2)^3}}\right. \\
		&\quad + \left.\frac{\pi_{++i}}{\sqrt{1-\sum_{l=1}^r\pi_{l++}^2}\sqrt{(1-\sum_{l=1}^r\pi_{++l}^2)^3}}\right), \\
		B_{ijk}^{\mathrm{miM}^*_1} &= \frac{\partial (\mathrm{miM}^*_1)}{\partial (\pi_{ijk})} \\
		&= \left\{\frac{-\pi_{++i}-\pi_{k++}}{\sqrt{1-\sum_{l=1}^r{\pi_{++l}^2}}\sqrt{1-\sum_{l=1}^r{\pi_{l++}^2}}} + (\sum_{l=1}^r\pi_{l+l}-\sum_{l=1}^r \pi_{l++}\pi_{++l})\left(\frac{\pi_{i++}}{\sqrt{1-\sum_{l=1}^r \pi_{++l}^2}\sqrt{(1-\sum_{l=1}^r\pi_{l++}^2)^3}}\right.\right. \\
		&\quad + \left.\left.\frac{\pi_{++k}}{\sqrt{1-\sum_{l=1}^r\pi_{l++}^2}\sqrt{(1-\sum_{l=1}^r\pi_{++l}^2)^3}}\right)\right\}, \\
		&i,j,k=1,\ldots,r;i\neq k,\\
		\bm{B}_{\mathrm{miM}^*_1}&=\frac{\partial(\mathrm{miM}^*_1)}{\partial (\bm{\pi}^{(3)})}=(B_{111}^{\mathrm{miM}^*_1}, B_{121}^{\mathrm{miM}^*_1},\ldots, B_{rrr}^{\mathrm{miM}^*_1})^\mathsf{T}.
	\end{align*}
	Furthermore,
	\begin{align*}
		B_{jii}^{\mathrm{miM}^*_2} &= \frac{\partial (\mathrm{miM}^*_2)}{\partial (\pi_{jii})} \\
		&= \frac{1-\pi_{+i+}-\pi_{++i}}{\sqrt{1-\sum_{l=1}^r{\pi_{++l}^2}}\sqrt{1-\sum_{l=1}^r{\pi_{+l+}^2}}} + (\sum_{l=1}^r \pi_{+ll}-\sum_{l=
			1}^r\pi_{+l+}\pi_{++l})\left(\frac{\pi_{+i+}}{\sqrt{1-\sum_{l=1}^r\pi_{++l}^2}\sqrt{(1-\sum_{l=1}^r\pi_{+l+}^2)^3}}\right. \\
		&\quad + \left.\frac{\pi_{++i}}{\sqrt{1-\sum_{l=1}^r\pi_{+l+}^2}\sqrt{(1-\sum_{l=1}^r\pi_{++l}^2)^3}}\right), \\
		B_{ijk}^{\mathrm{miM}^*_2} &= \frac{\partial (\mathrm{miM}^*_2)}{\partial (\pi_{ijk})} \\
		&= \left\{\frac{-\pi_{++i}-\pi_{+k+}}{\sqrt{1-\sum_{l=1}^r{\pi_{++l}^2}}\sqrt{1-\sum_{l=1}^r{\pi_{+l+}^2}}}+ (\sum_{l=1}^r\pi_{+ll}-\sum_{l=1}^r \pi_{+l+}\pi_{++l})\left(\frac{\pi_{+i+}}{\sqrt{1-\sum_{l=1}^r \pi_{++l}^2}\sqrt{(1-\sum_{l=1}^r\pi_{+l+}^2)^3}}\right.\right. \\
		&\quad + \left.\left.\frac{\pi_{++k}}{\sqrt{1-\sum_{l=1}^r\pi_{+l+}^2}\sqrt{(1-\sum_{l=1}^r\pi_{++l}^2)^3}}\right)\right\}, \\
		&i,j,k=1,\ldots,r;j\neq k,\\
		\bm{B}_{\mathrm{miM}^*_2}&=\frac{\partial(\mathrm{miM}^*_2)}{\partial (\bm{\pi}^{(3)})}=(B_{111}^{\mathrm{miM}^*_2}, B_{121}^{\mathrm{miM}^*_2},\ldots, B_{rrr}^{\mathrm{miM}^*_2})^\mathsf{T}.
	\end{align*}
	\begin{align*}
		\left[\frac{\partial(\bm{\mathrm{M}_3})}{\partial (\bm{\pi}^{(3)})}\right]^\mathsf{T}\bm{\phi}(\bm{\pi}^{(3)})\left[\frac{\partial(\bm{\mathrm{M}_3})}{\partial (\bm{\pi}^{(3)})}\right]=\begin{pmatrix}
			X_{\mathrm{miM}^*}& Z_{\mathrm{miM}^*} \\
			Z_{\mathrm{miM}^*}& Y_{\mathrm{miM}^*}
		\end{pmatrix},
	\end{align*}
	where
	\begin{align*}
		X_{\mathrm{miM}^*} &=\sum_{i=1}^{r}\sum_{j=1}^{r}\sum_{k=1}^{r}\pi_{ijk} (B_{ijk}^{\mathrm{miM}^*_1} )^2-\left\{\sum_{i=1}^{r}\sum_{j=1}^{r}\sum_{k=1}^{r}\pi_{ijk} B_{ijk}^{\mathrm{miM}^*_1} \right\}^2,\\
		Y_{\mathrm{miM}^*} &=\sum_{i=1}^{r}\sum_{j=1}^{r}\sum_{k=1}^{r}\pi_{ijk} (B_{ijk}^{\mathrm{miM}^*_2} )^2-\left\{\sum_{i=1}^{r}\sum_{j=1}^{r}\sum_{k=1}^{r}\pi_{ijk} B_{ijk}^{\mathrm{miM}^*_2} \right\}^2,\\
		Z_{\mathrm{miM}^*} &=\sum_{i=1}^{r}\sum_{j=1}^{r}\sum_{k=1}^{r}\pi_{ijk} B_{ijk}^{\mathrm{miM}^*_1}B_{ijk}^{\mathrm{miM}^*_2} -\left\{\sum_{i=1}^{r}\sum_{j=1}^{r}\sum_{k=1}^{r}\pi_{ijk} B_{ijk}^{\mathrm{miM}^*_1} \sum_{i=1}^{r}\sum_{j=1}^{r}\sum_{k=1}^{r}\pi_{ijk}B_{ijk}^{\mathrm{miM}^*_2} \right\}.
	\end{align*}
	Therefore, the asymptotic variance of $(\widehat{\mathrm{miM}^*}_1-\widehat{\mathrm{miM}^*}_2)$ is
	\begin{align*}
		(X_{\mathrm{miM}^*}+Y_{\mathrm{miM^*}}-2Z_{\mathrm{miM^*}})/n.
	\end{align*}
	
	\subsection*{Asymptotic variance with modified transformation method}
	\begin{align*}
		\sqrt{n}\{g \circ \psi_{l}( \hat{\bm{\pi}}^{(3)} ) - g \circ \psi_{l}( \bm{\pi}^{(3)} )\} &\xrightarrow{d} N \left(0, \{\nabla g \circ \psi_l(  \bm{\pi}^{(3)})\}^\mathsf{T}\phi(\bm{\pi}^{(3)})\{\nabla g \circ\psi_l(  \bm{\pi}^{(3)})\} \right).
	\end{align*}
	The derivation of the asymptotic variance for the transformed states using the modified transformation method can be performed based on the previously derived asymptotic variances of $\psi_{l}( \bm{\pi}^{(3)} )$.
	\[
	\nabla g\circ \psi_{l}( \bm{\pi}^{(3)} ) = g'(\psi_{l}( \bm{\pi}^{(3)} ))\nabla \psi_{l}( \bm{\pi}^{(3)} )=  \frac{2}{4-\psi_{l}( \bm{\pi}^{(3)} )^2}\nabla \psi_{l}( \bm{\pi}^{(3)} ).
	\]
	
\end{document}